# Ion transport and precipitation kinetics as key aspects of stress generation on pore walls induced by salt crystallization

A.Naillon[1,2], P.Joseph[2], M.Prat[1]*

[1]*Institut de Mécanique des Fluides de Toulouse (IMFT) - Université de Toulouse, CNRS-INPT-UPS, Toulouse, France*
[2]*LAAS-CNRS, Université de Toulouse, CNRS, Toulouse, France*

**Abstract**
The stress generation on pore walls due to the growth of a sodium chloride crystal in a confined aqueous solution is studied from evaporation experiments in microfluidic channels in conjunction with numerical computations of crystal growth. The study indicates that the stress build-up on the pore walls as the result of the crystal growth is a highly transient process taking place over a very short period of time (in less than 1s in our experiments). The analysis makes clear that what matters for the stress generation is not the maximum supersaturation at the onset of the crystal growth but the supersaturation at the interface between the solution and the crystal when the latter is about to be confined between the pore walls. It is shown that the stress generation can be characterized with a simple stress diagram involving the pore aspect ratio and the Damkhöler number characterizing the competition between the precipitation reaction kinetics and the ion transport towards the growing crystal. This opens up the route for a better understanding of the damage of porous materials induced by salt crystallization, an important issue in earth sciences, reservoir engineering and civil engineering.

Salt crystallization in pores causes damage in porous materials, a major issue in relation with building durability and cultural heritage conservation [1-4], underground structures [5], road [6] and geotechnical engineering [7]. A better understanding of the associated stress is also important in relation with geomorphology [8], concrete science [9] or the surface heave phenomenon of granular materials [10]. The fact that a growing crystal can generate stress has been known for more than a century [11], [12]. The key concept for the analysis of the stress generation is the crystallization pressure $P_c$ [13-15]. Corrections to the original expression [14] taking account the water activities and the crystal size have been developed, e.g. [16], [17], so that the current expression for sufficiently large crystals of sodium chloride (>1µm) reads,

$$P_c = \frac{2RT}{V_m}\left(\ln S + \ln \frac{\gamma_\pm}{\gamma_{\pm,0}}\right), \qquad (1)$$

where $R$ is the ideal gas constant, $T$ is the temperature, $V_m$ is the molar volume of the solid phase forming the crystal ($V_m$ = 27.02 cm$^3$/mol for NaCl), $\gamma_\pm$ the ion mean activity coefficient. Index $_0$ refers to the reference state where the crystal is in equilibrium with the solution. The ratio $S = m/m_0$ is the supersaturation, where $m$ denotes the molality of the solution ($S$ = 1 when the crystal and the solution are in equilibrium).

However, the mechanisms of stress generation are not yet well understood, e.g.[18]. For instance, no damage is observed in the experiments with glass capillary tubes presented in [19-20] whereas a supersaturation as high as 1.6 is obtained. Application of Eq.(1) for such a supersaturation leads to $P_c$ = 160 MPa, well above the glass wall tensile strength (~ 40 MPa). Therefore, it is clear that the mere knowledge of Eq.(1) and the maximum supersaturation reached in the pores are not sufficient to predict if damage will occur. A major challenge is thus to predict the conditions leading to damage. Here we analyze the stress generation mechanism from evaporation experiments performed in glass-polydimethylsiloxane (PDMS) microchannels (Fig.1). The main objective is to construct a stress generation diagram for this simple geometry. Although considered as less harmful than sodium sulfate, e.g. [3], [4], the case of sodium chloride is studied. This salt is very common, simpler that sodium sulfate since its crystallization leads to a unique non-hydrated form (referred to as halite) under standard conditions and can also lead to major damage [21].

---
* Corresponding author : mprat@imft.fr, +33 (0)5 34 32 28 83

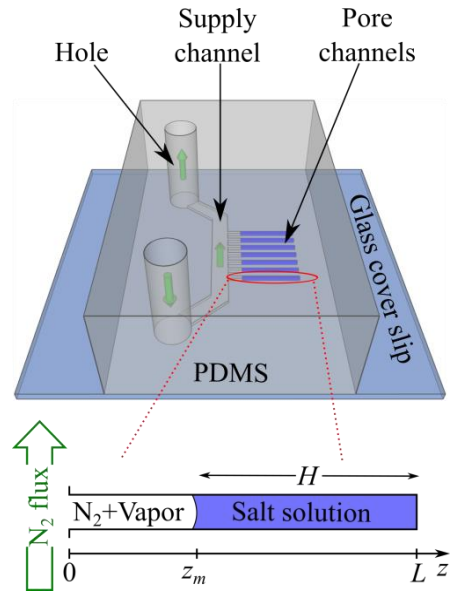

FIG. 1. (Color online) Schematic of the PDMS and glass microfluidic chip. Crystallization and wall deformations are observed in the pore channels.

As sketched in Fig.1, the experimental set-up is composed of a large channel which is used for supplying the fluids: salt solution or gaseous nitrogen. Smaller channels of 5×5 µm² square cross-section, referred to as pore channels, are positioned perpendicularly to the supply channel. Three pore channel lengths are tested: 100 µm, 200 µm and 300 µm. Two sodium chloride solution initial molalities are used: 1.89 and 4.25 mol/kg (the solubility is 6.15 mol/kg). This allows modifying the total amount of available salt in excess at the onset of crystallization. Salt with a purity ensured to be higher than 99.5% is dissolved in deionized water. Details on the microfluidic chip fabrication procedure are given in [22]. The crystallization is triggered by evaporation of the sodium chloride solution confined in the pore channels. Salt solution is provided from the top hole through the supply channel and invades the pore channels. Once the device is filled, a dry $N_2$ flux is imposed from the bottom hole to empty the supply channel and isolates salt solution in the pore channels. This flux is maintained during all the experiment to evaporate the solution. As a result of evaporation, the meniscus recedes into the pore channel, the ion concentration increases until the concentration $c_{cr}$ marking the onset of crystallization is reached. This leads to the formation of a single crystal, most often within the liquid bulk away from the receding meniscus. Then there is a rapid growth of the crystal within the channel. The supersaturation when crystal growth starts can be determined from a simple mass balance as explained in [22]. The supersaturation averaged over 99 experiments is 1.72, which is consistent with the values reported in previous works, [19], [20].

The experiments are performed at ambient temperature (22 to 24°C) on an inverted microscopy Zeiss Axio observer D1 working in transmission. Two video cameras are used: an Andor Zyla SCMos with a large field and a low frame rate (1 fps) to record the evaporation kinetics and the wall deformation at the end of the growth, and a high speed Photron Fastcam SA3 camera to record the rapid initial period of the crystal growth (1000 fps).

Movies are exploited thanks to the ImageJ© and Matlab© softwares to analyze the crystal growth by tracking the different interfaces (liquid-gas, crystal-liquid, crystal-pore wall).

Using a microfluidic PDMS device to analyze deformation due to sodium chloride crystallization is not a novelty [23]. However, contrary to [23], the size of our channels enables us to reproduce the situation of in-pore drying and to isolate sufficiently small volumes so that only one nucleation event occurs [24]. As shown in [22] and further illustrated here, our device is adapted for tracking precisely the crystal growth and for analyzing the evolution of the ion concentration around the crystal during its growth.

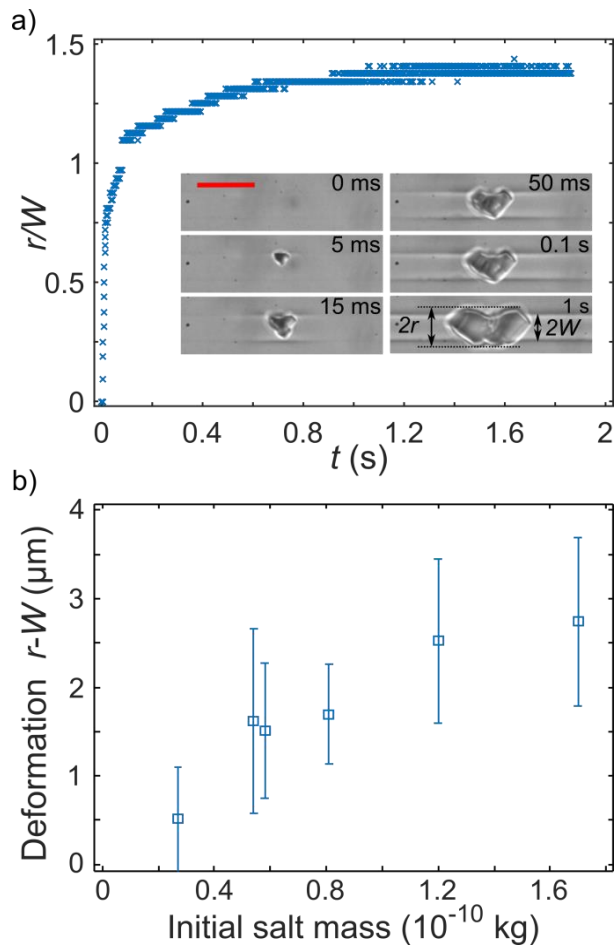

FIG. 2. (Color on line) Lateral crystal growth: a) Kinetics of crystal growth; $r$ is the crystal lateral half size, $W$ is the initial channel half width; red scale bar represents 10 µm. b) Absolute deformation against the initial dissolved salt mass.

As can be seen from Fig.2, a noticeable channel deformation is obtained and the growth is rapid (the channel walls are deformed in less than 1s). The maximum pore channel deformations (defined as the difference between the crystal half width $r$ and the initial channel half width $W$) range between 0 and 4 µm depending on the initial dissolved salt mass (equal to the pore channel volume times the initial concentration). As depicted in Fig.2b, the higher the initial salt mass, the higher the deformation is.

As reported in Supplemental Material [25], numerical simulations assuming purely elastic deformations and a uniform normal stress applied to the channel wall shows that a pressure of about 0.5 MPa is sufficient to obtain a deformation about equal to the maximum deformation observed in the experiment.

According to Eq.(1), this corresponds to a supersaturation of only 1.005, much lower than the supersaturation at the crystallization onset ($S \sim 1.7$). At first glance, this is surprising. Since the growth is fast (~1s), the change in the average concentration in the solution during the growth is expected to be small (the molecular diffusion of ions in the solution is $D_s \sim 10^{-9}$ m$^2$/s; with a liquid plug length $H$ of the order of 100 μm, a characteristic time of diffusion is $t = H^2/D_s = 10$s). However, what matters for the computation of the crystallization pressure from Eq.(1) is not the average supersaturation in the plug but the supersaturation at the interface between the crystal and the solution. The deformation computation results suggest that the supersaturation is in fact quite weak (i.e. slightly greater than 1) in the vicinity of the crystal when the latter is about to touch the wall on both opposite sides of the channel. To confirm this crucial point, the evolution of the ion concentration within the solution during the crystal growth must be analyzed. This is performed from numerical simulations using a model based on the diffusion reaction theory (DRT) [26]. First, crystal growth starts only once a stable nucleus appears in the metastable solution. Then the DRT distinguishes two steps: the transport of the ions from the solution to the crystal surface, followed by a reaction process during which ions fit in the crystal lattice. The latter is expressed as

$$w_{cr} = \frac{k_R}{\rho_{cr}}(c_i - c_{eq}) \qquad (2)$$

where $w_{cr}$ is the velocity of the crystal-solution interface; $k_R$ (m/s) is the reaction (precipitation) coefficient, $c_i$ (kg/m$^3$) is the salt mass concentration at the crystal surface, $c_{eq}$ is the mass concentration at equilibrium and $\rho_{cr}$ is the crystal density (kg/m$^3$).

Actually, $c_i$ is an unknown decreasing during the growth from the value $c_{cr}$ at the crystallization onset. The variation of $c_i$ results from the competition between the ion transport phenomena within the solution and the precipitation reaction. To obtain $c_i$ during the crystal growth and in particular when the crystal is about to reach the pore wall, the equations governing the ion transport within the solution during the crystal growth are solved

$$\frac{\partial \rho_l}{\partial t} + \nabla \cdot \rho_l \boldsymbol{v}_l = 0 \qquad (3)$$

$$\rho_l \left[ \frac{\partial \boldsymbol{v}_l}{\partial t} + \boldsymbol{v}_l \nabla \cdot \boldsymbol{v}_l \right] = -\nabla P_l + \mu_l \nabla^2 \boldsymbol{v}_l \qquad (4)$$

$$\frac{\partial \rho_l \omega_s}{\partial t} + \nabla \cdot (\rho_l \omega_s \boldsymbol{v}_l) = \nabla \cdot (\rho_l D_s \nabla \omega_s) \qquad (5)$$

where $\omega_s$ is the mass fraction of the ions in the solution, $\rho_l$ is the solution density, $\boldsymbol{v}_l$ is the solution velocity, $P_l$ is the pressure in the solution, $\mu_l$ is the solution dynamic viscosity, $t$ is the time. As can be seen from Eq.(5), the ion transport is governed by diffusion and convection since a velocity field is induced in the solution as a result of the crystal growth. The problem expressed by Eq.(3-5) together with the appropriate boundary conditions (given in [22]) is solved numerically using the commercial software COMSOL Multiphysics® for the simplified axisymmetric situation sketched in Fig. 3a. The supersaturation is everywhere 1.7 in the solution when the crystal growth computation starts (t = 0) and the computations are performed for different values of $W$, $H/W$ and $k_R$ around the experimental ones, respectively from 0.5 to 2.5 µm, 7 to 60 and $10^{-5}$ to $10^{-3}$ m/s. The growth being fast, the mass loss due to evaporation during the growth phase is neglected.

The initial shape of the crystal is modeled as a sphere for facilitating the numerical computations. Numerical tests have shown that the initial shape has little impact on the results.

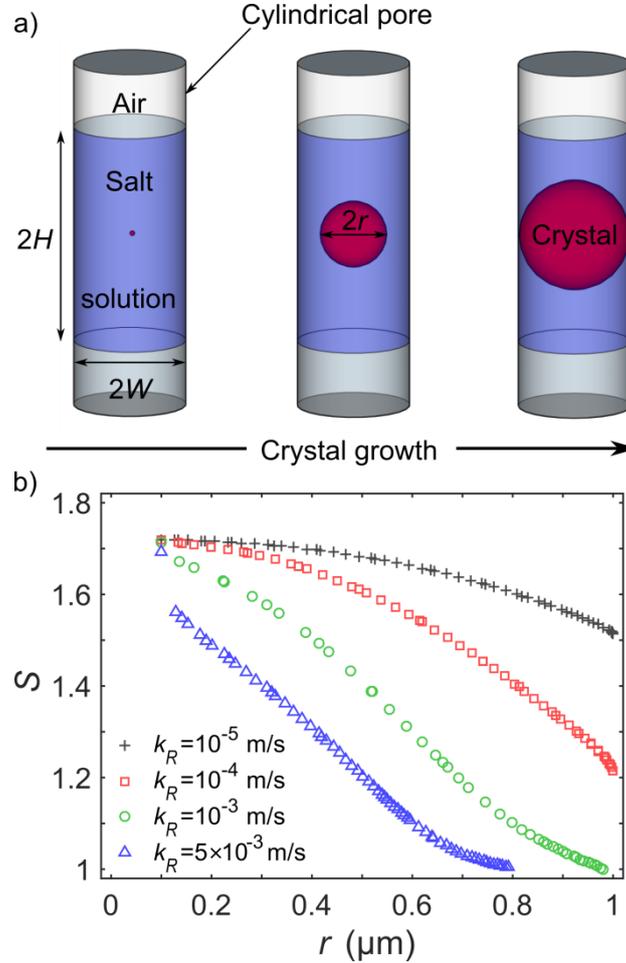

FIG. 3. (Color on line) Numerical simulations of crystal growth: a) Sketch of simulated problem. b) Supersaturation at the point of crystal surface located the closest to the wall during crystal growth for different $k_R$ ($W$=1 µm and $H$=60 µm).

Fig.3b explains why the stress generated on the wall is much smaller than the naive prediction based on the estimate using the supersaturation at the onset of crystallization. During the growth, there is a rapid decrease (in a few tens of ms) in the supersaturation at the crystal surface as the crystal develops in the solution. This result is highly dependent on the values of the reaction coefficient $k_R$. As discussed in [22], the growth rates determined from previous experiments in the literature are not representative of the sole reaction kinetics but are mostly controlled by the transport of the ions towards the growing crystal. As a result, the crystal growth rates reported in the literature are smaller than the intrinsic growth rate $k_R$ due to the precipitation reaction only. If $k_R$ is wrongly confused with the growth rate determined in the literature, i.e. $k_R \sim 10^{-5} - 10^{-4}$ m/s, the stress at the wall is much higher than the stress level necessary to cause the observed deformation. This is because the growth is very fast in the period controlled by the reaction. The analysis of our data suggests that $k_R$ is at least on the order of $2.3\times10^{-3}$ m/s [22].

As shown in Fig.3b, using a value of $k_R$ on the order of $10^{-3}$ m/s leads to a weak supersaturation (i.e. $S \sim 1$), compatible with the low stress required to obtain the observed deformation of PDMS in our experiment [25-§C]. The next step is to understand how the stress generation is related to the parameters of the problem, i.e. can we develop a stress generation diagram for our simple system?

To this end, the numerical model is used to determine the value of supersaturation when the crystal reaches the wall. The simulation is stopped when the crystal is 5 nm away from the pore wall, to be consistent with the presence of a few nanometers trapped liquid films [17], [27] (this film is necessary to supply ions to the growing crystal surface and can also transmit the stress between the crystal and the wall when it is sufficiently thin [27]). The corresponding supersaturation is referred to as the contact supersaturation. The contact crystallization pressure can then be evaluated from the contact supersaturation using Eq.(1).

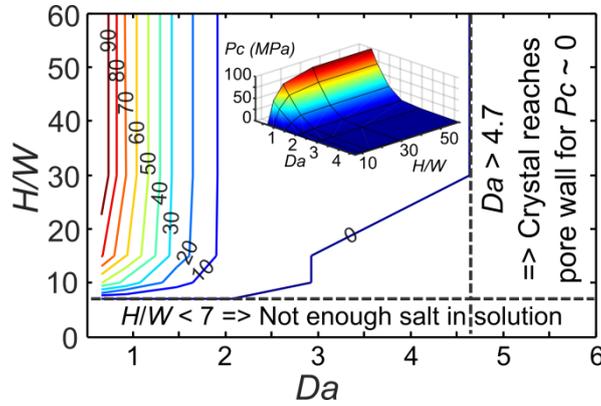

FIG. 4. (Color online) "Stress" generation diagram: computed values of contact crystallization pressure $P_c$ as a function of Damkhöler number $Da$, and channel aspect ratio $H/W$. Colored lines are isolines of $P_c$ with values indicated in MPa. The inset shows a 3D representation.

As illustrated in Fig.4, the resulting stress diagram depends on two parameters the aspect ratio $H/W$ and the Damkhöler number $Da = \sqrt{\frac{\rho_{cr} k_R}{\Delta c D_s}} W$, where $\Delta c = c_{cr} - c_{eq}$. First, no stress is generated when the crystal cannot reach the pore wall because there is not enough dissolved salt in excess in the plug at the crystallization onset, i.e. when $H/W \leq 3\Delta c/2\rho_{cr} \sim 7$ [25]. On the contrary, the contact crystallization pressure saturates for sufficiently high values of $H/W$ because the liquid plug is sufficiently long to behave as an infinite domain. Thus, the salt in excess far from the crystal is not consumed in the growth. Between these two limits, the contact supersaturation increases with increasing aspect ratio $H/W$.

The Damkhöler number $Da$ characterizes the competition between the precipitation reaction and the ion transport [26]: $Da=k_R/k_D$ where $k_D = \sqrt{D_s/t}$ characterizes the average ion mass transfer by diffusion toward the growing crystal after a time $t$. Taking as characteristic time the reaction time $t_R = W/w_{cr} \sim W\rho_{cr}/k_R\Delta c$ (time for the crystal to reach the wall when the reaction is limiting) leads to $Da = \sqrt{\frac{\rho_{cr} k_R}{\Delta c D_s}} W$. Simulation shows that this expression of $Da$ is relevant to characterize the crystal growth phenomenon. Simulations made for the same $Da$ but different $k_R$ and channel width $W$ lead to the same value of $S$ with a relative difference lower than 0.1%. Fig. 4 makes clear that $Da$ must be sufficiently small for a significant stress to be generated on pore walls. In practice, this means that the pores must be sufficiently small and explains why no mechanical damage was observed in the experiments reported in [19], [20]. A simple constraint on *Da* for stress generation can be expressed as follows. Supersaturation at crystal surface can remain high during the growth only if the diffusion rate is faster than the precipitation kinetics. The amount of salt needed to form a crystal of radius *W* can be estimated as $L_D W^2 \Delta c \sim W^3 \rho_{cr}$, where $L_D$ is the maximum distance over which ions are transported to form the crystal. The typical time to diffuse over a length $L_D$ is $t_D \sim L_D^2/D_s$, whereas the crystal reaches the pore wall after the reaction time $t_R$ when the reaction is the limiting process. Thus, a sufficient condition to observe a high contact supersaturation is $t_D \ll t_R$ or in dimensionless form $Da^2 \ll 1$, i.e.,

$$Da = \sqrt{\frac{D_s}{t_R k_R}} = \sqrt{\frac{\rho_{cr} k_R}{\Delta c D_s}} W \ll 1 \,. \qquad (6)$$

Varying both parameters *Da* and *H/W*, the contact supersaturation varies between 1 and 1.33 for *Da* > 0.65 and *H/W* > 7. As depicted in Fig.4, this corresponds to a crystallization pressure varying between 0 and 98 MPa using Eq.(1) (to be compared to the tensile strength of sedimentary rocks, which is on the order of 1-10 MPa [28]).

In summary the analysis presented in this letter makes clear that what matters for the stress generation is not the maximum saturation at the onset of the crystal growth but the supersaturation at the interface between the solution and the crystal when the latter is about to be confined between the pore walls. The generation of stresses on the pore walls as the result of the growth of a sodium chloride single crystal is actually a highly transient nonequilibrium process occurring over a very short period (in less than 1s in our experiments). This is because the precipitation reaction is quite fast. As a result, the supersaturation at the crystal interface rapidly decreases during its growth. Note also that this process eventually leads to a permanent deformation (see the Supplemental Material [25-§E] for more details). This better understanding of the stress generation mechanisms enables us to propose a simple stress diagram for a single pore involving the pore aspect ratio *H/W* and the Damkhöler number. It is surmised that this opens up the route for diagrams for more complex geometry such as the pore space of a porous medium (as briefly discussed in the Supplemental Material [25-§F]). It must also be noted that the fact that the crystal growth is quite fast makes challenging to model the stress generation process within the framework of the classical continuum approach to porous media because this type of approach is typically not well adapted for accounting for rapid events at pore scale.

Acknowledgements: Financial supports from ANDRA, CNRS-INSIS and CNRS NEEDS-MIPOR program are gratefully acknowledged. This work was partly supported by LAAS-CNRS micro and nano technologies platform member of the French RENATECH network.

[1]   A. S. Goudies and H. A. Viles, Salt Weathering Hazards. (Wiley, Chichester, 1997).
[2]   R. M. Espinosa-Marzal and G. W. Scherer, Acc. Chem. Res. **43**, 897 (2010).
[3]   R. J. Flatt, F. Caruso, A. M. A. Sanchez, and G. W. Scherer, Nat. Commun. **5**, 4823 (2014).
[4]   M. Schiro, E. Ruiz-Agudo, and C. Rodriguez-Navarro, Phys. Rev. Lett. **109**, 265503 (2012).
[5]   L. Oldecop and E. Alonso, Int. J. Rock Mech. Min. Sci. **54**, 90 (2012).
[6]   B. Obika, R. J. Freer-Hewish, and P. G. Fookes, Q. J. Eng. Geol. Hydrogeol. **22**, 59 (1989).
[7]   H. W. Wellman and A. T. Wilson, Nature **205**, 1097 (1965)

[8] K. Serafeimidis and G. Anagnostou, Rock Mech. Rock Eng. **46**, 619 (2013).
[9] E. E. Alonso and A. Ramon, Géotechnique **63**, 857 (2013).
[10] R. Hird and M. D. Bolton, Proc. R. Soc. A **472**, (2016).
[11] J. Lavalle, Compt. rend. hebdo. Acad. Sci. **36,** 493 (1853).
[12] G. F. Becker and A. L. Day, Proc. Wash. Acad. Sci. **7**, 283 (1905).
[13] S. Taber, Am. J. Sci. **4**, 532 (1916).
[14] C. W. Correns, Discuss. Faraday Soc. **5**, 267 (1949).
[15] L.A. Rijniers, H.P. Huinink, L. Pel, and K. Kopinga, Phys. Rev. Lett. **94**, 075503 (2005).
[16] G. W. Scherer, Cem. Concr. Res. **34**, 1613 (2004).
[17] M. Steiger, J. Cryst. Growth **282**, 455 (2005).
[18] O. Coussy, J. Mech. Phys. Solids **54**, 1517 (2006).
[19] A. Naillon, P. Duru, M. Marcoux, and M. Prat, J. Cryst. Growth **422**, 52 (2015).
[20] J. Desarnaud, H. Derluyn, J. Carmeliet, D. Bonn, and N. Shahidzadeh, J. Phys. Chem. Lett. **5**, 890 (2014).
[21] C. Noiriel, F. Renard, M.-L. Doan, and J.-P. Gratier, Chem. Geol. **269**, 197 (2010).
[22] A. Naillon, P. Joseph, and M. Prat, J. Cryst. Growth **463**, 201 (2017).
[23] K. Sekine, A. Okamoto, and K. Hayashi, Am. Mineral. **96**, 1012 (2011).
[24] R. Grossier and S. Veesler, Cryst. Growth Des. **9**, 1917 (2009).
[25] See Supplemental Material at [url will be inserted by publisher] for more details on deformation computation and determination of the geometrical critical ratio.
[26] J. W. Mullin, *Crystallization*, (Butterworth Heinemann, Oxford, 2001), 4$^{th}$ edition.
[27] J. Desarnaud, D. Bonn, and N. Shahidzadeh, Sci. Rep. **6**, 30856 (2016).
[28] M. A. Perras and M. S. Diederichs, Geotech. Geol. Eng., **32**, 525 (2014).